\documentclass[12pt]{article}
\usepackage{authblk}

\begin{document}

\title{\ Nonlinear Ordinary Differential Equations: A discussion on
Symmetries and Singularities}
\author[1]{Andronikos Paliathanasis\thanks{%
anpaliat@phys.uoa.gr}}
\author[2,3,4]{ PGL Leach\thanks{%
leach@ucy.ac.cy}}
\small{
\affil[1]{\small{Instituto de Ciencias F\'{\i}sicas y Matem\'{a}ticas, Universidad
Austral de Chile, Valdivia, Chile}}
\affil[2]{Department of Mathematics and
Institute of Systems Science, Research and Postgraduate Support, Durban
University of Technology, PO Box 1334, Durban 4000, Republic of South Africa}
\affil[3]{School of Mathematics, Statistics and Computer Science, University
of KwaZulu-Natal, Private Bag X54001, Durban 4000, Republic of South Africa}
\affil[4]{Department of Mathematics and Statistics, University of Cyprus,
Lefkosia 1678, Cyprus}}
\maketitle

\begin{abstract}
Two essential methods, the symmetry analysis and of the singularity analysis, for the study of the integrability of nonlinear ordinary differential equations are discussed. The main similarities and differences of these two different methods are given.
\end{abstract}

\section{ Introduction}

{ The systematic analysis of the symmetries and singularities of
ordinary differential equations began in the last quarter of the nineteenth
century. In a series of papers and books \cite{Lie 1, Lie 2, Lie 3, Lie 4,
Lie 5, Lie 6}, seeking to do for ordinary differential equations what Galois
had done for algebraic equations, the Norwegian mathematician, Sophus Lie,
wrought a much greater achievement which even to this day influences every
area in which differential equations, indeed difference equations, arise.
The genius of Lie's work was to take the infinitesimal representations of
the finite transformations of continuous groups, thereby moving from the
group to a local algebraic representation, and to study the invariance
properties under them. This resulted in linearization of all equations
and/or functions under consideration. The infinitesimal transformation
\begin{equation}
\bar{x}=x+\varepsilon \xi ,\quad \quad \bar{y}=y+\varepsilon \eta ,
\label{1.1}
\end{equation}%
where $\varepsilon $ is the infinitesimal parameter of the transformation,
could be represented in terms of the differential operator $G$ given by
\begin{equation}
G=\xi \frac{\partial u}{\partial x}+\eta \frac{\partial u}{\partial y}
\label{1.2}
\end{equation}%
as the deformation from the identity
\begin{equation}
\bar{x}=\left( 1+\varepsilon G\right) x,\quad \quad \bar{y}=\left(
1+\varepsilon G\right) y.  \label{1.3}
\end{equation}%
The effect of the infinitesimal transformation on functions or equations
involving derivatives could be determined by the extension of $G$ to deal
with derivatives. For the first derivative we have
\begin{eqnarray}
\frac{\bar{y}}{\bar{x}} &=&\frac{y+\varepsilon \eta }{x+\varepsilon \xi }
\nonumber \\
&=&\frac{y^{\prime }+\varepsilon \eta ^{\prime }}{1+\varepsilon \xi ^{\prime
}}  \nonumber \\
&=&y^{\prime }+\varepsilon (\eta ^{\prime }-y^{\prime }\xi ^{\prime })
\label{1.4}
\end{eqnarray}%
and we write the first extension of $G$ as
\begin{equation}
G^{[1]}=G+(\eta ^{\prime }-y^{\prime }\xi ^{\prime })\partial _{y^{\prime }}.
\label{1.5}
\end{equation}%
The extension for higher derivatives is determined in the same fashion and
we can write the $n$th extension in a recursive form as \cite{Mahomed 90}
\begin{equation}
G^{[n]}=G^{[n-1]}+\left\{ \eta ^{(n)}-\sum_{i=0}^{n-1}\left(
\begin{array}{c}
n \\
i+1%
\end{array}%
\right) y^{(n-i)}\xi ^{(i+1)}\right\} \partial _{y^{(n)}}.  \label{1.6}
\end{equation}
}

{ A function or differential equation containing up to the $n$th
derivative invariant under the action of $G^{[n]}$ is said to possess the
symmetry $G$. This is expressed as
\begin{equation}
G^{[n]}f(x,y,y^{\prime (n)})=0  \label{1.6.1}
\end{equation}%
in the case of the function $f(x,y,y^{\prime (n)})$ and
\begin{equation}
G^{[n]}f(x,y,y^{\prime (n)})_{\left\vert _{f=0}\right. }=0  \label{1.6.2}
\end{equation}%
in the case of the $n$th-order differential equation, $f(x,y,y^{\prime
(n)})=0$. The set of all such symmetries constitutes a Lie algebra under the
operation of taking the Lie Bracket, namely
\begin{equation}
\left[ G_{i},G_{j}\right] _{LB}=G_{i}G_{j}-G_{j}G_{i}.  \label{1.7}
\end{equation}%
In the case of point symmetries, \textit{ie}, the coefficient functions $\xi
$ and $\eta $ are functions of $x$ and $y$ only, the algebraic properties of
the set of symmetries ${G_{i},i=1,m},$ are invariant under extension. In the
case of contact symmetries, for which the coefficient functions can also
contain $y^{\prime }$ in such a way that
\begin{equation}
\frac{\partial \eta }{\partial y^{\prime }}=y^{\prime }\frac{\partial \xi }{%
\partial y^{\prime }},  \label{1.8}
\end{equation}%
the same applies. When one has ensured that the first extension has a
coefficient depending upon $x$, $y$ and $y^{\prime }$ only, the algebraic
properties must be established using the set ${G_{i}^{[1]},i=1,m}$. In the
case of generalized symmetries, for which the dependence upon derivatives is
limited only by the order of the differential equation in the case that a
differential equation is being considered, and in the case of nonlocal
symmetries, in which the coefficient functions can depend upon integrals in
a nontrivial fashion, the calculation of the algebraic properties is a
somewhat more complicated affair. Although generalised and nonlocal
symmetries play important roles in certain problems, fortunately the
algebraic difficulties have, to date, not been a great problem in the study
of these types of symmetries of differential equations. }

{ The original work of Lie was motivated by geometric considerations
and he commenced with point transformations, and so point symmetries, and
then extended his work to include contact symmetries so that the
transformations were from $(x,y)$ space to $(\bar{x},\bar{y})$ space or from
$(x,y,y^{\prime })$ space to $(\bar{x},\bar{y},\bar{y^{\prime }})$ space.
The use of generalized transformations was firmly established by the work of
Noether on the invariance of the Action Integral of the Calculus of
Variations under infinitesimal transformations \cite{Noether 18}. The use of
nonlocal symmetries arose in the last part of the twentieth century. }

{ The major thrust of the singularity analysis of differential
equations is associated with the French school led by Painlev\'{e} in the
last years of the nineteenth century and the early years of the twentieth
century \cite{Painleve 1, Painleve 2, Painleve 3, Painleve 4} following its
successful application to the determination of the third integrable case of
Euler's equations for a spinning top by Kowalevskaya \cite{Kowalevski 88}.
Since then considerable work has been done on the classification problem of
higher-order and higher-degree ordinary differential equations by Bureau
\cite{Bureau 1, Bureau 2, Bureau 3} and Cosgrove \textit{et al} \cite%
{Cosgrove 1,Cosgrove 2, Cosgrove 3}. Application to partial differential
equations became widespread in the second half of the twentieth century with
significant contributions being made by Kruskal \cite{Kruskal 62 a}. The
development of the Painlev\'{e} Test for the determination of integrability
of a given equation or system of equations and its systematization in the
ARS algorithm \cite{Ablowitz 78 a,Ablowitz 80 a, Ablowitz 80 b} has made the
singularity analysis a routine tool for the practising applied
mathematician. Popular expositions such as that found in the review by
Ramani, Grammaticos and Bountis \cite{Ramani 89} and the monograph of Tabor
\cite{Tabor 89} provide clear guides for the implementation of the
algorithm. More precise prescriptions are found in the somewhat more
technical works of Conte \cite{Conte 94 a, Conte 99 a}. }

{ The basic purpose of both forms of analysis is to facilitate the
solution of differential equations. The existence for a differential
equation of a sufficient number of Lie symmetries of the right type enables
one to solve the differential equation by means of repeated reduction of
order and a reverse series of quadratures or by means of the determination
of a sufficient number of first integrals. The latter is the route taken in
Noether's Theorem. In singularity analysis the differential equation (or
system of differential equations; this should be implied unless specifically
excluded) is deemed integrable if it possesses the Painlev\'e Property. The
Painlev\'e Property in brief is that the differential equation possesses a
Laurent expansion about a movable polelike singularity in the complex plane
of the independent variable with the requisite number of arbitrary constants
to provide the general solution of differential equation. The Laurent
expansion implies that the solution of the differential equation is analytic
except at the singularities. This is somewhat stronger than that which the
Lie approach gives and there still exists the question of a complete
reconciliation between the two approaches. For a differential equation of
more than simple complexity there is the possibility of different patterns
of singular behaviour and it is the conventional wisdom \cite{Tabor 89} [p
300] that for each pattern of singular behaviour there must exist a solution
in terms of an analytic function, expressed in terms of a Laurent expansion,
with the requisite number of arbitrary constants for the equation to possess
the Painlev\'e Property and so be integrable. (In all of these
considerations we must allow for the possibility of the so-called `weak
Painlev\'e Property' in which the polelike singularity is replaced by an
algebraic branch point and the Laurent expansion is in terms of fractional
powers of the complex variable. The discussion is \textit{mutatis mutandis}
identical.) However, this very convenient criterion has been shown, by way
of counterexample, to be too strong \cite{Leach 2000}. }

{ In the context of the Lie analysis the concept of integrability is
not as strong as it is in the Painlev\'e analysis. For the latter the
solution of the differential equation must be analytic apart from isolated
movable polelike singularities or have branch point singularities so that
the complex plane of the independent variable can be divided into sections
and in each section the solution is analytic. One can say that a
differential equation is integrable in the sense of Lie if it possesses a
sufficient number of symmetries for it to be reducible to an algebraic
equation, although generally one would not bother with the ultimate step and
be content with reduction to a separable first-order ordinary differential
equation, \textit{ie} to a quadrature. To the sufficiency of the number of
symmetries one must sound a note of caution. If an $n $th-order system has $%
n $ Lie point symmetries with a solvable algebra, one knows that the system
is reducible to quadratures. The absence of this property does not
immediately obviate the possibility of integrability. In the process of
reduction of order of a system using the known symmetries additional point
symmetries may arise. Such symmetries are known as Type II hidden symmetries
\cite{Abraham-Shrauner 1, Abraham-Shrauner 2, Abraham-Shrauner 3} and
originate from nonlocal symmetries in the main. A similar phenomenon can be
observed when the order of a differential equation is increased, a technique
occasionally of use in the solution of certain equations \cite{Govinder 1,
Govinder 2}. A Lie point symmetry which arises on the increase of order of a
differential equation is called a Type I hidden symmetry and originates in a
nonlocal symmetry of the original equation. }

{ The possession of the Painlev\'{e} Property is representation
dependent and its preservation under transformation is guaranteed only in
the case of a M\"{o}bius transformation. A Lie symmetry exists independently
of the representation. In fact one could say that symmetries can be neither
created nor destroyed. The particular nature of a symmetry can change with
the representation. Thus the origin of hidden symmetries, \textit{ie} Lie
point symmetries which appear, as it were, from nowhere on a change of order
of the differential equation, is found in nonlocal symmetries\footnote{%
Occasionally contact or generalise symmetries, but generally nonlocal
symmetries.}. The symmetry was there all the time, but was hidden from view
due to the restriction to the viewing of point symmetries only. Preservation
of the type of Lie symmetry is guaranteed only by a transformation of the
same quality as that of the symmetry. Point symmetries are preserved under
point transformations, contact symmetries are preserved under contact
transformations, generalized symmetries are preserved under generalized
transformations and nonlocal symmetries under nonlocal transformations. }

{ In addition to their use in the solution of differential equations
Lie symmetries are used for the classification of equations and the
establishment of equivalence classes of equations, \textit{ie} those
equations obtainable by means of transformations of the same nature as the
symmetries used in the classification. Usually the classification is in
terms of Lie point symmetries. In the case of third-order equations contact
symmetries are also used. Because the Lie point symmetries constitute an
algebra under the operation of taking the Lie Bracket, the classification is
conveniently commenced from the algebra. The task of establishing all
representations of Lie algebras which admit a differential equation of given
order is a tediously formidable task. The reader is referred to the works of
Mubarakzyanov \cite{Mubar 1,Mubar 2,Mubar 3} for the classification scheme
of Lie algebras and of the Montreal School for the representations \cite%
{Sharp 1, Sharp 2, Sharp 3}. Additional to these systematic investigations
there are results for specific types of equations. There have also been
studies of the algebraic properties of first integrals of differential
equations \cite{Abraham-Shrauner 4, Govinder 3, Flessas 97 a}. More recently
the concept of a complete symmetry group, \textit{ie} the number of
symmetries required to specify completely a differential equation or its
first integrals, has attracted attention \cite{Krause 94, Nucci 96,
Andriopoulos 01 a} and has been used to show the identity of a number of
nonlinear systems of somewhat different properties \cite{Nucci 01}. }

{ Before we commence our treatment of nonlinear ordinary differential
equations we present a short summary of the results for linear equations.
The Painlev\'e analysis is not really relevant to linear systems as there is
no question of their possessing movable singularities. However, linear
systems are known to be integrable. A scalar $n\geq 3 $th-order linear
equation has either $n+ 1 $, $n+ 2 $ or $n+ 4 $ Lie point symmetries. In the
case of a linear second-order equation the number of Lie point symmetries is
always eight. Apart from the symmetry related to the very linearity of the
equation all other symmetries require a knowledge of the solution of the
original equation and so the theoretical plethora of Lie point symmetries is
of no great practical value. Systems of linear equations have been little
studied. In the case of systems of two second-order linear equations the
number of symmetries has been shown to range from 5 to 15 \cite%
{Gonsalez-Gascon 84, Mahomed 90} and in the case of $n $ autonomous
second-order linear systems to range from $n+ 1 $ to $(n+ 2) ^ 2-1 $ with
the latter being a representation of the algebra $sl (n+ 2,R) $ \cite%
{Gorringe 87}. In the case of $n $ equations of the $m $th order the maximum
number of Lie point symmetries is $n ^ 2+nm + 3 $ \cite{Nucci 01b}. }

\section{ Applications of Lie symmetries to nonlinear ordinary
differential equations}

{ The calculation of the Lie symmetries of a function or differential
equation, be it ordinary or partial, linear or nonlinear, is a tedious
business except for the simplest of expressions. One is advised to make use
of the codes written in one of the symbolic manipulation packages, such as
Nucci's interactive code \cite{Nucci 90, Nucci 96b}, the more automated code
of Head \cite{Head 93, Head 97} or the Mathematica Add-on, Sym, developed by
Dimas \cite{Dimas 05 a, Dimas 06 a, Dimas 08 a, Andriopoulos 09 a}. The
codes of Head, Nucci and Dimas are readily available. These codes are
designed primarily for the computation of Lie point symmetries, but they can
be used for the computation of contact and generalized symmetries and some
specialised forms of symmetry such as approximate \cite{Moyo 00}. There is
an application for which these codes are not so satisfactory and that is the
calculation of possibly nonlocal symmetries to determine the existence of
integrating factors \cite{Bouquet 01}. }

{ It is possible that a nonlinear ordinary differential equation is in
fact a linear ordinary differential equation written using inappropriate
variables. This in fact is the case with many of the nonlinear equations
listed in the classic compendium of Kamke \cite{Kamke 92}. We select one of
them, \cite{Kamke 92} [6.51, p 554], namely
\begin{equation}
y^{\prime \prime }+f(y)y^{\prime 2}+g(x)y^{\prime }=0.  \label{2.1}
\end{equation}%
A second-order equation requires two Lie point symmetries to be reducible to
quadratures. Given that an equation with which we are working may not be
written in the ideal variables, we must allow for the possibility that one
of the symmetries could be nonlocal. For a second-order equation the
determining equation, (\ref{1.6.2}), becomes
\begin{equation}
\frac{\partial f}{\partial y}+\left( \eta ^{\prime }-y^{\prime }\xi ^{\prime
}\right) \frac{\partial f}{\partial y^{\prime }}+\left( \eta ^{\prime \prime
}-2y^{\prime \prime }\xi ^{\prime }-y^{\prime }\xi ^{\prime \prime }\right)
\frac{\partial f}{\partial y^{\prime \prime }}  \label{2.2}
\end{equation}%
For the calculation of nonlocal symmetries (\ref{2.2}) is extremely awkward.
One can regard it as a second-order linear equation in either $\xi $ or $%
\eta $ with the other function being at one's disposal. It is usually
convenient to be regarded as a linear equation in $\eta $ and to set $\xi =0$%
. Provided that one can solve the equation for $\eta $, there are two
symmetries, as required. However, there are four possible two-dimensional
Lie algebras with a standard representation and it is necessary to find the
equivalent form in the representation we have adopted to solve (\ref{2.2}).
We list these in Table \ref{table 1} (adapted from \cite{Bouquet 01} [Table
1]. }

\begin{center}
{
\begin{equation}
\begin{tabular}{|c|c||l|l|l|}
\hline
$\mbox {\rm Type}$ & $\left\{ G_1,G_2\right\}$ & $\mbox {\rm Canonical forms}
$ & $\mbox {\rm Form of}$ & $\mbox {\rm Present forms}$ \\
&  & $\mbox {\rm of $G_1 $ and $G_2 $}$ & $\mbox {\rm equation}$ & $%
\mbox
{\rm of $G_1 $ and $G_2 $}$ \\ \hline
$\mbox {\rm I}$ & 0 & $G_1 =\frac{\partial u}{\partial x}$ & $y^{\prime
\prime }=f\left(y^{\prime }\right)$ & $G_1 =y^{\prime }\frac{\partial u}{%
\partial y}$ \\
&  & $G_2 =\frac{\partial u}{\partial y}$ &  & $G_2 =\frac{\partial u}{%
\partial y}$ \\
$\mbox {\rm II}$ & 0 & $G_1 =\frac{\partial u}{\partial y}$ & $y^{\prime
\prime }=f (x)$ & $G_1 =\frac{\partial u}{\partial y}$ \\
&  & $G_2 =x\frac{\partial u}{\partial y}$ &  & $G_2 =x\frac{\partial u}{%
\partial y}$ \\
$\mbox {\rm III}$ & $G_1$ & $G_1 =\frac{\partial u}{\partial y}$ & $%
xy^{\prime \prime }=f \left(y^{\prime }\right)$ & $G_1 =\frac{\partial u}{%
\partial y}$ \\
&  & $G_2 =x\frac{\partial u}{\partial x} +y\frac{\partial u}{\partial y}$ &
& $G_2 =\left(xy^{\prime }-y\right)\frac{\partial u}{\partial y}$ \\
$\mbox {\rm IV}$ & $G_1$ & $G_1 =\frac{\partial u}{\partial y}$ & $y^{\prime
\prime }=y^{\prime }f (x)$ & $G_1 =\frac{\partial u}{\partial y}$ \\
&  & $G_2 =y\frac{\partial u}{\partial y}$ &  & $G_2 =y\frac{\partial u}{%
\partial y}$ \\ \hline
\end{tabular}
\label{table 1}
\end{equation}
}
\end{center}

{ We note that in cases Type II and Type IV the normal form of the
equation is of the form we desire to use. }

{ When we apply this procedure to (\ref{2.1}), we obtain the equation
\begin{equation}
\left( \frac{\eta ^{\prime }}{y^{\prime }}\right) ^{\prime }+\left( \eta
f\right) ^{\prime }=0  \label{2.3}
\end{equation}%
which has a fairly obvious solution and so we have the two symmetries and
Lie Bracket
\begin{eqnarray}
G_{1} &=&\exp \left[ -\int f(y)y^{\prime }\mbox{\rm d}x\right] \frac{%
\partial u}{\partial y}, \\
G_{2} &=&\exp \left[ -\int f(y)y^{\prime }\mbox{\rm d}x\right] \int
y^{\prime }\exp \left[ \int f(y)y^{\prime }\mbox{\rm d}x\right] \mbox{\rm d}x%
\frac{\partial u}{\partial y}\frac{\partial u}{\partial y},
\end{eqnarray}%
\begin{equation}
\left[ G_{1},G_{2}\right] _{LB}=G_{1}
\end{equation}%
so that our second-order equation, (\ref{2.1}), is of Lie's Type IV and is
transparently linear when written in the variables
\begin{equation}
x=x,\quad w=\exp \left[ \int f(y)y^{\prime }\mbox{\rm d}x\right] \quad
\mbox
{\rm as}\quad w^{\prime \prime }+g(x)w^{\prime }=0  \label{2.5}
\end{equation}%
the integration of which is theoretically quite trivial. }

{ In general we may apply one of the codes to determine the Lie point
symmetries of a given differential equation. For example the well-known
equation \cite{Golubev 50, Moreira 84, Ervin 84, Chisholm 87}
\begin{equation}
y^{\prime \prime }+3yy^{\prime 3}=0,  \label{2.6}
\end{equation}%
often called the Painlev\'{e}-Ince Equation, has the eight Lie point
symmetries \cite{Mahomed 85}
\begin{eqnarray}
&&G_{1}=\mbox{$\frac{1}{2}$}x^{2}y\frac{\partial u}{\partial x}+\left(
xy^{2}-\mbox{$\frac{1}{2}$}x^{2}y^{3}-y\right) \frac{\partial u}{\partial y}
\nonumber \\
&&G_{2}=y\frac{\partial u}{\partial x}-y^{3}\frac{\partial u}{\partial y}
\nonumber \\
&&G_{3}=xy\frac{\partial u}{\partial x}+\left( y^{2}-xy^{3}\right) \frac{%
\partial u}{\partial y}  \nonumber \\
&&G_{4}=\left( -\mbox{$\frac{1}{2}$}x^{2}y+x\right) \frac{\partial u}{%
\partial x}+\left( \mbox{$\frac{1}{2}$}x^{2}y^{3}-xy^{2}\right) \frac{%
\partial u}{\partial y}  \nonumber \\
&&G_{5}=\left( \mbox{$\frac{1}{3}$}x^{3}-\mbox{$\frac{1}{4}$}x^{4}y\right)
\frac{\partial u}{\partial x}+\left( -x-x^{3}y^{2}+\mbox{$\frac{1}{4}$}%
x^{4}y^{3}+\mbox{$\frac{3}{4}$}x^{2}y\right) \frac{\partial u}{\partial y}
\nonumber \\
&&G_{6}=\left( -\mbox{$\frac{1}{2}$}x^{3}y+x^{2}\right) \frac{\partial u}{%
\partial x}+\left( xy+\mbox{$\frac{1}{2}$}x^{3}y^{3}-\mbox{$\frac{3}{2}$}%
x^{2}y^{2}\right) \frac{\partial u}{\partial y}  \nonumber \\
&&G_{7}=\left( -\mbox{$\frac{1}{2}$}x^{3}y+\mbox{$\frac{3}{2}$}x^{2}\right)
\frac{\partial u}{\partial x}+\left( 1+\mbox{$\frac{1}{2}$}x^{3}y^{3}-%
\mbox{$\frac{3}{2}$}x^{2}y^{2}\right) \frac{\partial u}{\partial y}
\nonumber \\
&&G_{8}=\frac{\partial u}{\partial x}  \label{2.7}
\end{eqnarray}%
and so is equivalent under a point transformation to the equation
\begin{equation}
\frac{\mbox{\rm d}^{2}Y}{\mbox{\rm d}X^{2}}=0\Rightarrow Y=AX+B,\quad
A,B\quad \mbox {\rm constants}.  \label{2.8}
\end{equation}%
The point transformation is
\begin{equation}
Y=-\mbox{$\frac{1}{2}$}x^{2}+\frac{x}{y},\quad X=x-\frac{1}{y}  \label{2.9}
\end{equation}%
and so the solution of (\ref{2.6}) is obviously
\begin{equation}
y=\frac{2(1+Ax)}{Ax^{2}+2x+C}.  \label{2.10}
\end{equation}%
In the more general form,
\begin{equation}
y^{\prime \prime }+kyy^{\prime 3}=0,  \label{2.11}
\end{equation}%
the equation has only the two obvious symmetries of invariance under
translation in $x$ and rescaling, namely
\begin{equation}
G_{1}=\frac{\partial }{\partial x}\quad \mbox {\rm and}\quad G_{2}=-x\frac{%
\partial }{\partial x}+y\frac{\partial }{\partial y}  \label{2.12}
\end{equation}%
and the solution is given by following expression
\begin{eqnarray}
y\left( x\right) &=&k^{-\frac{1}{3}}\left( 3\left( x+y_{1}\right) +k^{-\frac{%
1}{2}}\left( 8y_{0}^{3}+9\left( x+y_{1}\right) ^{2}k\right) \right) ^{\frac{1%
}{3}}+  \label{2.13} \\
&&-2y_{0}\left( 3\left( x+y_{1}\right) +k^{-\frac{1}{2}}\left(
8y_{0}^{3}+9\left( x+y_{1}\right) ^{2}k\right) \right) ^{-\frac{1}{3}}
\nonumber
\end{eqnarray}%
where $y_{0},~y_{1}$ are the constant of integration. The performance of the
quadrature in closed form is generally not possible and the inversion to
obtain $y(x)$ even less so \cite{Lemmer 93}. We meet (\ref{2.6}) and (\ref%
{2.11}) below in the discussion of the Painlev\'{e} Property. }

{ In the case of third-order equations there is the possibility of
linearization by means of a point transformation or by means of a contact
transformation. The need to distinguish between the two possibilities arises
from some differences in algebraic properties between second-order and
third-order linear equations. In general second-order equations can have 0,
1,2,3 or 8 Lie point symmetries. Third-order linear equations can have 4, 5
or 7 Lie point symmetries and in the last case there are 10 Lie contact
symmetries. Although a third-order linear equation cannot have six Lie point
symmetries, there is no such restriction on a third-order nonlinear
equation. For example the Kummer-Schwarz Equation
\begin{equation}
2y^{\prime }y^{\prime \prime \prime }-3y^{\prime \prime 2}=0  \label{2.14}
\end{equation}%
has the 6 Lie point symmetries
\begin{eqnarray}
&G_{1}=\displaystyle{\frac{\partial u}{\partial x}}&G_{4}=\frac{\partial u}{%
\partial y}  \nonumber \\
&G_{2}=x\displaystyle{\frac{\partial u}{\partial x}}&G_{5}=y\frac{\partial u%
}{\partial y}  \nonumber \\
&G_{3}=x^{2}\displaystyle{\frac{\partial u}{\partial x}}\quad &G_{6}=y^{2}%
\frac{\partial u}{\partial y}  \label{2.15}
\end{eqnarray}%
with the Lie algebra $sl(2,R)\oplus sl(2,R)$. It does, moreover, have 10 Lie
contact symmetries and so can be transformed to the third-order equation of
maximal symmetry by means of a contact transformation
\begin{equation}
\frac{\mbox{\rm d}^{3}Y}{\mbox{\rm d}X^{3}}=0.  \label{2.16}
\end{equation}%
By way of interest the generalized Kummer-Schwarz Equation
\begin{equation}
y^{\prime }y^{\prime \prime \prime }+ny^{\prime \prime 2}=0,  \label{2.17}
\end{equation}%
which possesses only the four Lie point symmetries $G_{1}$, $G_{2}$, $G_{4}$
and $G_{5}$ of the six listed in (\ref{2.15}), is also equivalent to the
equation of (\ref{2.16}) but by means of the nonlocal transformation,
\begin{equation}
X=x,\quad Y=\int y^{\prime n+1}\mbox{\rm d}x,  \label{2.18}
\end{equation}%
corresponding to the local symmetry $G_{1}$ of (\ref{2.17}) rather than the
contact transformation of (\ref{2.16}). }

\section{Singularity analysis of nonlinear ordinary
differential equations}

{ In the spirit of Sophie Kowalevskaya \cite{Kowalevski 88}, we seek
to determine whether or not a given differential equation possesses movable
singularities. To take an example consider the Painleve-Ince Equation, }

{
\begin{equation}
y^{\prime \prime }+3yy^{\prime }+y^{3}=0,  \label{3.01}
\end{equation}%
which has attracted a certain amount of attention in recent decades \cite%
{Abraham-Shrauner 1,Bou,Ervin 84,Mahomed 85,Leach88a,Lemmer 93}. If a
movable singularity exists then the solution of the latter equation will
described by the power-law function $y\left( x\right) \simeq \left(
x-x_{0}\right) ^{p}$, where $p$ is a negative number and $x_{0}$ indicates
the position of the singularity. Of a movable singularity because the value
of $x_{0}$, the position, depends on the initial conditions, that is,
different initial conditions provide us with different positions for the
singular point. }

{ We substitute $y\left( x\right) =a_{0}\left( x-x_{0}\right)
^{p}=a_{0}\chi ^{p}$ in (\ref{3.01}) and obtain the expression
\begin{equation}
a_{0}p\left( p-1\right) \chi ^{p-2}+3p\left( a_{0}\right) ^{2}\chi
^{2p-1}+a_{0}\chi ^{3p},  \label{3.02}
\end{equation}%
for which balance occurs if $p=-1$ and consequently $a_{0}=1$ or $a_{0}=2$.
This means that the movable singularity is a simple pole and there are two
possibilities which follow from the leading-order behaviour. The arbitrary
location of the movable singularity gives one of the constants of
integrations. As (\ref{3.01}) is a second-order equation,\ the other
constant of integration has to be determined from a series developed about
the singularity. }

{ We select the leading order $a_{0}=1$ write the solution as a
Laurent expansion of the form%
\begin{equation}
y\left( x\right) =\chi ^{-1}+Y\left( \chi \right)  \label{3.03}
\end{equation}%
where $Y\left( \chi \right) $ represents the remained of the Laurent
expansion. To avoid the tedium of coping with an infinite series we simply
replace $Y\left( \chi \right) $ with $\mu \chi ^{-1+s}$ in (\ref{3.01}) and
obtain
\begin{equation}
\mu \chi ^{-s}\left( s-1\right) \left( s+1\right) +3\mu ^{2}s\chi ^{2s}+\mu
^{3}\chi ^{3s}=0.  \label{3.04}
\end{equation}
}

{ We take the term linear in $\mu $ and equate it to zero so that $s=-1
$,~$s=1$. The former value $s=-1$ is to be expected as it is associated with
the movable singularity. The second value $s=1$, indicates the term in the
series at which the second constant of integration occurs. We infer that the
series increments by integral powers and write the Laurent expansion as
\begin{equation}
y\left( x\right) =\chi ^{-1}+\sum\limits_{I=1}^{\infty }a_{I}\chi ^{-1+I}.
\label{3.05}
\end{equation}
}

{ When we substitute (\ref{3.05}) into (\ref{3.01}), we obtain a
recurrence relation for the remaining coefficients. \ \ The first terms are $%
a_{2}=-a_{1}^{2}$, $a_{3}=a_{1}^{3}$,~$a_{4}=-a_{1}^{4}$,~$a_{5}=a_{1}^{5}$%
..., that is,
\begin{equation}
a_{I}=\left( -1\right) ^{-1+I}a_{1}^{I}.  \label{3.06}
\end{equation}
}

{ In the case that we consider the second leading order $a_{0}=2$, the
Laurent expansion that we find that it is a decreasing series. This is known
as a Left Painlev\'{e} Series in contrast to the former results which was a
Right Painlev\'{e} Series. }

{ The explanation for the two distinct types of solutions is simple.
We are in the complex plane and we integrate around the singularity. An
increasing expansion means that we integrate from the singularity until a
border. The decreasing series means that we integrate from the border to
infinity. However, there exists a possibility that the Laurent expansion
admits increasing and decreasing terms. The explanation of latter is that we
integrate over annulus around the singularity which has two borders \cite%
{Andriopoulos 06 a}. }

{ This approach to the singularity analysis has been succinctly
summarised in the papers of Ablowitz, Ramani and Segur \cite{Ablowitz 78 a,
Ablowitz 80 a, Ablowitz 80 b} and it is called the ARS algorithm from the
initials of the authors. That algorithm can be briefly described as follows:
}

\begin{itemize}
\item { Determine the leading-order behaviour, at least in terms of
dominated exponent. The coefficient of the leading-order term may or may not
be explicit. }

\item { Determine the exponents at which the arbitrary constants of
integration enter. }

\item { Substitute an expansion up to the maximum resonance into the
full equation to check for consistency. }
\end{itemize}

{ For the singularity analysis to work the exponents of the
leading-order term needs to be a negative integer or a nonintegral rational
number. \ Equally the resonances have to be rational numbers. Excluding the
generic resonance $s=-1$, for a Right Painlev\'{e} Series the resonances
must be nonnegative, for a Left Painlev\'{e} Series the resonances must be
nonpositive while for a full Laurent expansion the resonances they have to
be mixed. Clearly for a second-order ordinary equation \ the possible
Laurent expansions are Left or Right Painlev\'{e} Series. }

\section{ Symmetries and Singularities}

{ There are similarities and dissimilarities between the two methods,
approaches for determining the integrability of differential equations.
Singularity analysis indicates potential integrability. Symmetry analysis
can give stronger results in that it can provide a route to the explicit
solution of the equation in closed form. In the route of the latter there is
the question the choice of the group invariant transformation which leaves
invariant the differential equation. There is a choice of the type of
symmetry, such as point symmetries, contact symmetries, Lie-B\"{a}cklund
symmetries, nonlocal symmetries and many others. The ease of applicability
deteriorates with the increasing complexity of the functional forms permitted
in the coefficient functions. }

{ Singularity analysis is straighforward in principle as it does not
offer so many choices. However, singularity analysis is coordinate dependent
which is not true for symmetry analysis. To demonstrate that we give
two well-known elementary equations, the free particle and the
\textquotedblleft hyperbolic\textquotedblright\ oscillator. Both of these
equations are invariant under an eight-dimensional Lie algebra, $%
sl\left( 3,R\right) $, i.e., they are maximally symmetric. That means that
there exists a transformation which transform the one equation to the other
one and vice versa. }

{ From the singularity point of view, equations $y^{\prime \prime }=0$
and $y^{\prime \prime }-y=0$ do not possess any movable singularity and the
ARS algorithm that we discussed above fails. However, that does not mean
that there does not exist a coordinate system in which these two equations
pass the singularity test. For the equation of motion of the free particle
that is simple by selecting the new variable $w=y^{-1}$. Then the new
equation is%
\begin{equation}
ww^{\prime \prime }-2\left( w^{\prime }\right) ^{2}=0  \label{ss01}
\end{equation}%
which admits the leading term with exponent $p=-1$ and arbitrary $a_{0}$. Easily we have
that the resonances are $s=-1$, and $s=0$. The second value is expected
because the leading term has arbitrary constant, $a_{0}$, and a zero
resonance provides us with that property. }

{ As far as concerns the linear equation $y^{\prime \prime }=y$, which
does not pass the singularity test we perform the change of variables
\begin{equation}
x=-\ln \left( u\left( v\right) \right) ~,~y=\frac{du\left( v\right) }{dv}{, }
\end{equation}%
and we have the third-order nonlinear equation%
\begin{equation}
u^{2}u_{,v}u_{,vvv}+\left( u~\left( u_{v}\right) ^{2}-u^{2}\right)
u_{,vv}-4\left( u_{,v}\right) ^{4}=0
\end{equation}%
which passes the singularity test hence the integrability is expressed also in
terms of the singularity analysis. }

{One of the main differences between the two methods is that symmetries
provide for the conservation laws of a dynamical system, {\it ie} functions which are
invariant in time. This is generally not necessarily true in the case of singularity analysis. The
conservation laws which follow from the symmetry analysis are applied for
the analysis of the dynamical system as they provide surfaces in the
phase space in which the solution evolves. On the other hand the solution
which follows from the singularity analysis admits the correct number of
constants of integration but in general information about the nature of
conservation laws cannot be extracted. For instance for the free particle (%
\ref{ss01}) it is easy to extract the conservation law }$I=\frac{1}{w^{2}}%
\dot{w}$\ from the symmetry vector $w\partial _{w}$ {\ which
is nothing else than the law of conservation of momentum. However, from the
singularity analysis the solution of (\ref{ss01}) is given as a Laurent
expansion and one needs to calculate all of the coefficients to determine
the solution.  The conservation law cannot be determined, which
does not mean that the conservation law does not exist. }

{ The two forms of analysis can be regarded as complementary. Symmetry
analysis is very effective when it works, singularity analysis is also very
effective when it works. Neither method is a complete answer to question of
integrability for the simple reason that there exist equations for which
neither method provides a result, \ but which are trivially integrable.} \

{ In this paper we have concentrated upon nonlinear differential
equations for the purposes of clarity of presentation. The considerations
here can be extended \textit{mutatis mutandus} to systems of ordinary
differential equations and to partial differential equations. }

\section*{\protect\small Acknowledgements}

{ AP acknowledges financial support of FONDECYT grant no. 3160121. PGL
Leach thanks the Instituto de Ciencias F\'{\i}sicas y Matem\'{a}ticas of the
UACh for the hospitality provided while this work carried out and
acknowledges the National Research Foundation of South Africa and the
University of KwaZulu-Natal for financial support. The views expressed in
this work should not be attributed to either institution. \newline
}


\begin{thebibliography}{99}
\bibitem{Lie 1} { Sophus Lie \textit{Theorie der
Transformationsgrupprn: Vol I} (Chelsea, New York) (1970) }

\bibitem{Lie 2} { Sophus Lie \textit{Theorie der
Transformationsgrupprn: Vol II} (Chelsea, New York) (1970) }

\bibitem{Lie 3} { Sophus Lie \textit{Theorie der
Transformationsgrupprn: Vol III} (Chelsea, New York) (1970) }

\bibitem{Lie 4} { Sophus Lie \textit{Differentialgleichungen}
(Chelsea, New York) (1967) }

\bibitem{Lie 5} { Sophus Lie \textit{Continuerliche Gruppen} (Chelsea,
New York) (1971) }

\bibitem{Lie 6} { Sophus Lie \textit{Geometrie der
Berhrungstransformationen} (Chelsea, New York) (1977) }

\bibitem{Mahomed 90} { F. M. Mahomed \& P. G. L. Leach Symmetry Lie
algebras of $n$th order ordinary differential equations \textit{Journal of
Mathematical Analysis and Applications} \textbf{151} (1990) 80-107 }

\bibitem{Noether 18} { Emmy Noether Invariante Variationsprobleme
\textit{K\"oniglich Gesellschaft der Wissenschaften G\"ottingen Nachrichten
Mathematik-physik Klasse} \textbf{2} (1918) 235-267 }

\bibitem{Painleve 1} { Painlev\'e P, {Le\c{c}ons sur la th\'eorie
analytique des \'equations diff\'erentielles} (Le\c{c}ons de Stockholm,
1895) (Hermann, Paris, 1897). Reprinted, {O$\!$euvres de Paul Painlev\'e},
vol.~I, \'Editions du CNRS, Paris, 1973. }

\bibitem{Painleve 2} { Painlev\'e P (1900) M\'emoire sur les
\'equations diff\'erentielles du second ordre dont l'int\'egrale
g\'en\'erale est uniforme \textit{Bulletin of the Mathematical Society of
France} \textbf{28} 201-265 }

\bibitem{Painleve 3} { Painlev\'e P (1902) Sur les \'equations
diff\'erentielles du second ordre et d'ordre sup\'erieur dont l'int\'egrale
g\'en\'erale est uniforme \textit{Acta Mathematica} \textbf{25} 1-85 }

\bibitem{Painleve 4} { Painlev\'{e} P (1906) Sur les \'equations
diff\'erentielles du second ordre \`a points critiques fixes \textit{Comptes
Rendus de la Acad\'emie des Sciences de Paris} \textbf{143} 1111-1117 }

\bibitem{Kowalevski 88} { Kowalevski Sophie (1889) Sur la probl\`{e}me
de la rotation d'un corps solide autour d'un point fixe \textit{Acta Math}
\textbf{12} 177-232 }

\bibitem{Bureau 1} { Bureau FJ (1964) Differential equations with
fixed critical points \textit{Annali di Matematica pura ed applicata}
\textbf{LXIV} 229-364 }

\bibitem{Bureau 2} { Bureau FJ (1964) Differential equations with
fixed critical points \textit{Annali di Matematica pura ed applicata}
\textbf{LXVI} 1-116 }

\bibitem{Bureau 3} { Bureau FJ (1972) \'{E}quations diff\'{e}%
rentielles du second ordre en $Y$ et du second degr\'{e} en $\ddot{Y}$ dont
l'int\'{e}grale g\'{e}n\'{e}rale est \'{a} points critiques fixes \textit{%
Annali di Matematica pura ed applicata} \textbf{XCI} 163-281 }

\bibitem{Bureau 4} { Bureau FJ (1972) Integration of some nonlinear
systems of ordinary differential equations \textit{Annali di Matematica pura
ed applicata} \textbf{XCVI} 345-359 }

\bibitem{Cosgrove 1} { Cosgrove CM (2000) Higher-order Painlev\'e
equations in the polynomial class 1. Bureau Symbol P2 \textit{Studies in
Applied Mathematics} \textbf{104} 1-65 (DOI:10.1111/1467-9590.00130) }

\bibitem{Cosgrove 2} { Cosgrove CM (2000) Chazy classes IX-XI of
third-order differential equations \textit{Studies in Applied Mathematics}
\textbf{104} 171-228 (DOI:10.1111/1467-959000134) }

\bibitem{Cosgrove 3} { Cosgrove CM (2006) Higher-order Painlev\'e
equations in the polynomial class II. Bureau Symbol P1 \textit{Studies in
Applied Mathematics} \textbf{116} 321-413 (DOI:10.1111/j.1467-9590.00346.x) }

\bibitem{Kruskal 62 a} { Kruskal M (1962) Asymptotic theory of
Hamiltonian and other systems with all solutions nearly periodic \textit{%
Journal of Mathematical Physics} \textbf{3} 806-828 }

\bibitem{Ablowitz 78 a} { Ablowitz MJ, Ramani A \& Segur H (1978)
Nonlinear Evolution Equations and Ordinary Differential Equations of
Painlev\'e Type \textit{Lettere alNuovo Cimento} \textbf{23} 333-337 }

\bibitem{Ablowitz 80 a} { Ablowitz M J, Ramani A \& Segur H (1980) A
connection between nonlinear evolution equations and ordinary differential
equations of P Type I \textit{Journal of Mathematical Physics} \textbf{21}
715-721 }

\bibitem{Ablowitz 80 b} { Ablowitz M J, Ramani A \& Segur H (1980) A
connection between nonlinear evolution equations and ordinary differential
equations of P type II \textit{Journal of Mathematical Physics} \textbf{21}
1006-1015 }

\bibitem{Ramani 89} { Ramani A, Grammaticos B \& Bountis T The
Painlev\'e property and singularity analysis of integrable and nonintegrable
systems \textit{Physics Reports} \textbf{180} (1989) 159-245 }

\bibitem{Tabor 89} { Tabor M (1989) \textit{Chaos and Integrability in
Nonlinear Dynamics} (John Wiley, New York) }

\bibitem{Conte 94 a} { Conte R (1994) Singularities of differential
equations and integrability in \textit{Introduction to Methods of Complex
Analysis and Geometry for Classical Mechanics and Nonlinear Waves} Benest D
and Fr\oe schl\'e C edd (\'{E}ditions Fronti\`{e}res, Gif-sur-Yvette) 49-143
}

\bibitem{Conte 99 a} { Conte R (1999) \textit{The Painlev\'e Property:
One Century Later} Conte Robert ed (CRM Series in Mathematical Physics,
Springer-Verlag, New York) }

\bibitem{Leach 2000} { Leach PGL, Cotsakis S \& Flessas GP (2000)
Symmetry, singularity and integrability in complex dynamics II: Rescaling
and time-translation in two-dimensional systems \textit{Journal of
Mathematical Analysis and Application} \textbf{251} 587-608 }

\bibitem{Abraham-Shrauner 1} { Abraham-Shrauner B (1993) Hidden
symmetries and linearisation of the modified Painlev\'{e}-Ince equation
\textit{Journal of Mathematical Physics} \textbf{34} 4809-4816 }

\bibitem{Abraham-Shrauner 2} { Abraham-Shrauner B \& Guo A (1993)
Hidden and nonlocal symmetries of nonlinear differential equations in
\textit{Modern Group Analysis: Advanced Analytical and Computational Methods
in Mathematical Physics} Ibragimov NH, Torrissi M \& Valenti A edd (Kluwer,
Dordrecht) 1-5 }

\bibitem{Abraham-Shrauner 3} { Abraham-Shrauner B \& Leach PGL (1993)
Hidden symmetries of nonlinear ordinary differential equations in \textit{%
Exploiting Symmetry in Applied and Numerical Analysis} Allgower E, George K
\& Miranda R edd (Lecture Notes in Applied Mathematics \textbf{29} American
Mathematical Society, Providence) 1-10 }

\bibitem{Govinder 1} { Govinder KS, Athorne C \& Leach PGL (1993) The
algebraic structure of generalized Ermakov systems in three dimensions
\textit{Journal of Physics A: Mathematical and General} \textbf{26}
4035-4046 }

\bibitem{Govinder 2} { Govinder KS \& Leach PGL (1997) A group
theoretic approach to a class of second order ordinary differential
equations not possessing Lie point symmetries \textit{Journal of Physics A:
Mathematical and General} \textbf{30} 2055-2068 }

\bibitem{Mubar 1} { Mubarakzyanov GM (1963) On solvable Lie algebras
\textit{Izvestia Vysshikh Uchebn Zavendeni\u{\i} Matematika} \textbf{32}
114-123 }

\bibitem{Mubar 2} { Mubarakzyanov GM (1963) Classification of real
structures of five-dimensional Lie algebras \textit{Izvestia Vysshikh Uchebn
Zavendeni\u{\i} Matematika} \textbf{34} 99-106 }

\bibitem{Mubar 3} { Mubarakzyanov GM (1963) Classification of solvable
six-dimensional Lie algebras with one nilpotent base element \textit{%
Izvestia Vysshikh Uchebn Zavendeni\u{\i} Matematika} \textbf{35} 104-116 }

\bibitem{Sharp 1} { Patera J \& Winternitz P (1975) Continuous
subgroups of the fundamental groups of Physics. I. General method and the
Poincar\newline
'e group \textit{Journal of Mathematical Physics} \textbf{16} 1597-1614 }

\bibitem{Sharp 2} { Patera J, Sharp RT, Winternitz P \& Zassenhaus H
(1976) Invariance of real low dimension Lie algebras \textit{Journal of
Mathematical Physics} \textbf{17} 986-994 }

\bibitem{Sharp 3} { Patera J \& Winternitz P (1977) Algebras of real
three- and four-dimensional Lie algebras \textit{Journal of Mathematical
Physics} \textbf{18} 1449-1455 per }

\bibitem{Abraham-Shrauner 4} { Abraham-Shrauner B, Govinder KS \&
Leach PGL (1995) Integration of second order equations not possessing point
symmetries \textit{Physics Letters A} \textbf{203} 169-174 }

\bibitem{Govinder 3} { Govinder KS \& Leach PGL (1995) The algebraic
structure of the first integrals of third-order linear equations \textit{%
Journal of Mathematical Analysis and Applications} \textbf{193} 114-133 }

\bibitem{Flessas 97 a} { Flessas GP, Govinder KS \& Leach PGL (1997)
Characterisation of the algebraic properties of first integrals of scalar
ordinary differential equations of maximal symmetry \textit{Journal of
Mathematical Analysis and Applications} \textbf{212} 349-374 }

\bibitem{Krause 94} { Krause J (1994) On the complete symmetry group
of the classical Kepler system \textit{Journal of Mathematical Physics}
\textbf{35} 5734-5748 }

\bibitem{Nucci 96} { Nucci MC (1996a) Interactive REDUCE programs for
calculating Lie point, non-classical, Lie-B\"{a}cklund, and approximate
symmetries of differential equations: manual and floppy disk \textit{CRC
Handbook of Lie Group Analysis of Differential Equations. Vol. III: New
Trends}, Ibragimov NH ed (Boca Raton: CRC Press) }

\bibitem{Andriopoulos 01 a} { Andriopoulos K, Leach PGL \& Flessas GP
(2001) Complete symmetry groups of ordinary differential equations and their
integrals: some basic considerations \textit{Journal of Mathematical
Analysis and Application} \textbf{262} 256-273 }

\bibitem{Nucci 01} { Nucci MC \& Leach PGL (2001) The harmony of the
Kepler and related problems \textit{Journal of Mathematical Physics} \textbf{%
42} }

\bibitem{Gonsalez-Gascon 84} { Gonz\'alez-Gasc\'on F \&
Gonz\'alez-L\'opez A (1983) Symmetries of differential equations IV \textit{%
Journal of Mathematical Physics} \textbf{24} 2006-2021 }

\bibitem{Gorringe 87} { Gorringe VM \& Leach PGL (1988) Lie point
symmetries for systems of second order linear ordinary differential
equations \textit{Qu\ae stiones Mathematic\ae } \textbf{11} 95-117 }

\bibitem{Nucci 01b} { Nucci MC \& Leach PGL (2001) Maximal algebras of
Lie point symmetries for systems of ordinary differential equations
(preprint: Department of Mathematics and Informatics, University of Perugia,
06123 Perugia, Italy) }

\bibitem{Nucci 90} { Nucci MC (1990) Interactive REDUCE programs for
calcuating classical, non-classical and Lie-B\"acklund symmetries for
differential equations (preprint: Georgia Institute of Technology, Math
062090-051) }

\bibitem{Nucci 96b} { Nucci MC (1996) The complete Kepler group can be
derived by Lie group analysis \textit{Journal of Mathematical Physics}
\textbf{37} 1772-1775 }

\bibitem{Head 93} { Head AK (1993) \texttt{LIE}, a PC program for Lie
analysis of differential equations \textit{Computer Physics Communications}
\textbf{77} 241-248 }

\bibitem{Head 97} { Sherring J, Head AK \& Prince GE (1997) Dimsym and
LIE: symmetry determining packages \textit{Mathematical and Computer
Modelling} \textbf{25} 153-164 }

\bibitem{Andriopoulos 09 a} { Andriopoulos K, Dimas S, Leach PGL \&
Tsoubelis D (2009) On the systematic approach to the classification of
differential equations by group theoretical methods \textit{Journal of
Computational and Applied Mathematics} \textbf{230} 224-232 (DOI:
10.1016/j.cam.2008.11.002) }

\bibitem{Dimas 05 a} { Dimas S \& Tsoubelis D (2005) SYM: A new
symmetry-finding package for Mathematica \textit{Group Analysis of
Differential Equations} (Ibragimov NH, Sophocleous C \& Damianou PA edd
University of Cyprus, Nicosia) 64--70 }

\bibitem{Dimas 06 a} { Dimas S \& Tsoubelis D (2006) A new
Mathematica-based program for solving overdetermined systems of PDEs.
\textit{8th International Mathematica Symposium}. Avignon, France. }

\bibitem{Dimas 08 a} { Dimas S (2008) \textit{Partial Differential
Equations: Algebraic Computing and Nonlinear Systems} (PhD thesis,
University of Patras, Patras, Greece) }

\bibitem{Moyo 00} { Leach PGL, Moyo S, Cotsakis S \& Lemmer RL (2001)
Symmetry, singularities and integrability in complex dynamics III:
Approximate symmetries and invariants \textit{Journal of Nonlinear
Mathematical Physics} \textbf{8} 139-156 }

\bibitem{Bouquet 01} { Leach PGL \& Bouquet S\'E (2002) Symmetries and
integrating factors \textit{Journal of Nonlinear Mathematical Physics}
\textbf{9} \textbf{Second Supplement} 73-91 }

\bibitem{Kamke 92} { Kamke E (1983) \textit{Differentialgleichungen
L\"osungsmethoden und L\"osungen} (BG Teubner, Stuttgart) }

\bibitem{Golubev 50} { Golubev VV (1950) \textit{Lectures on
Analytical Theory of Differential Equations} (Gosteekhizdat,
Moscow-Leningrad) }

\bibitem{Moreira 84} { Moreira IC, A Note on the Invariants for the
Time-Dependent Oscillator, \textit{J. Phys. A: Math. Gen.} \textbf{18}
(1985), 899--907. }

\bibitem{Ervin 84} { Ervin VJ, Ames WF \& Adams E (1984) Nonlinear
waves in pellet fusion in \textit{Wave Phenomena: Modern Theory and
Applications} Rodgers C \& Moodie TB edd (North Holland, Amsterdam) }

\bibitem{Chisholm 87} { Chisholm JSR \& Common AK (1987) a class of
second-order differential equations and related first-order systems \textit{%
Journal of Physics A: Mathematical and General} \textbf{20} 5459-5472 }

\bibitem{Mahomed 85} { Mahomed FM \& Leach PGL (1985) The linear
symmetries of a nonlinear differential equation \textit{Qu\ae stiones
Mathematic\ae } \textbf{8} 241-274 }

\bibitem{Lemmer 93} { Lemmer RL \& Leach PGL (1993) The Painlev\newline
'e test, hidden symmetries and the equation $y^{\prime \prime }+yy^{\prime
3}=CO$ \textit{Journal of Physics A: Mathematical and General} \textbf{26}
5017-5024 }

\bibitem{Bou} { Bouquet S\'{E}, Feix MR \& Leach PGL (1991) Properties
of second-order ordinary differential equations invariant under
time-translation and self-similar transformation, Journal of Mathematical
Physics \textbf{32} 1480-1490 }

\bibitem{Leach88a} { Leach PGL, Feix MR \& Bouquet S\'{E} (1988)
Analysis and solution of a nonlinear second-order differential equation
through rescaling and through a dynamical point of view \textit{Journal of
Mathematical Physics} \textbf{29} 2563-2569 }

\bibitem{Andriopoulos 06 a} { Andriopoulos K \& Leach PGL (2006) An
interpretation of the presence of both positive and negative nongeneric
resonances in the singularity analysis \textit{Physics Letters A} \textbf{359%
} 199-203 }

\bibitem{Andriopoulos 08 a} { Andriopoulos K \& Leach PGL (2008) The
Mixmaster Universe: The final reckoning? \textit{Journal of Physics A:
Mathematical and Theoretical} \textbf{41} 155201 (11 pages)
(DOI:10.1088/1751-8113/41/15/155202) }

\bibitem{Emden 07} { Emden R (1987) \textit{Gaskugeln, Anwedungen der
mechanishen Warmentheorie auf Kosmologie und meteorologische Probleme} (BG
Teubner,Leipzig) }

\bibitem{Fowler 14} { Fowler RH (1914) The form the infinity of real,
continuous solutions of a certain differential equation of the second order
\textit{Quart J Math } \textbf{45} 289-350 }

\bibitem{Hereman 96 a} { Hereman W (1996) \textit{CRC Handbook of Lie
Group Analysis of Differential Equations Vol. 3: New Trends} (CRC Press,
Boca Raton) Chapter 13
}

\bibitem{Krause 95 a} { Krause J (1995) On the complete symmetry group
of the Kepler problem in Arima A ed \textit{Proceedings of the XXth
International Colloquium on Group Theoretical Methods in Physics} (World
Scientific, Singapore) 286-290 }

\bibitem{Kruskal 62 a} { Kruskal M (1962) Asymptotic theory of
Hamiltonian and other systems with all solutions nearly periodic \textit{%
Journal of Mathematical Physics} \textbf{3} 806-828 }

\bibitem{Lane 69} { Lane IJ Homer (1869-70) On the theoretical
temperature of the Sun under the hypothesis of a gaseous mass maintaining
its volume by its internal heat and depending on the laws of gases known to
terrestrial experiment \textit{American Journal of Science \& Arts} \textbf{4%
} 57-74 }

\bibitem{Thompson 60} { Thompson W (Lord Kelvin) (1860-1862) On the
convective equilibrium of temperature in the atmosphere \textit{Manchester
Philosophical Society Proceedings} \textbf{2} 170-176 }
\end{thebibliography}
\end{document}